\documentclass[conference]{IEEEtran}
\IEEEoverridecommandlockouts
\usepackage{cite}
\usepackage{adjustbox}
\usepackage{amsmath,amssymb,amsfonts}
\usepackage{graphicx}
\usepackage{textcomp}
\usepackage{xcolor}
\usepackage{multicol}
\usepackage{multirow}
\usepackage{subcaption}
\usepackage{balance}
\usepackage{enumitem}
\usepackage{booktabs}
\usepackage{array}
\newcolumntype{x}[1]{>{\centering\arraybackslash\hspace{0pt}}p{#1}}

\usepackage{pst-all}
\usepackage{pstricks-add}
\usepackage{float}

\usepackage{pgfplots}

\usepackage{tikz}
\usepackage{tikzscale}
\usepackage{transparent}

\usepackage{algorithm}
\usepackage[noend]{algpseudocode}

\usepackage{comment}

\usepackage{blindtext}

\definecolor{purplelight}{HTML}{ead1dc}
\definecolor{purpledark}{HTML}{a64d79}

\definecolor{purple}{HTML}{733d87}

\definecolor{color1}{HTML}{e6194B}

\definecolor{color2}{HTML}{3cb44b}

\definecolor{color3}{HTML}{ffe119}

\definecolor{color4}{HTML}{4363d8}

\definecolor{color5}{HTML}{f58231}

\definecolor{color6}{HTML}{911eb4}

\definecolor{color7}{HTML}{42d4f4}

\definecolor{color8}{HTML}{f032e6}

\definecolor{color9}{HTML}{bfef45}

\definecolor{color10}{HTML}{fabebe}

\definecolor{color11}{HTML}{469990
}
\definecolor{color12}{HTML}{e6beff}

\definecolor{color13}{HTML}{9A6324}

\definecolor{color14}{HTML}{a9a9a9}

\definecolor{color15}{HTML}{800000}

\def\BibTeX{{\rm B\kern-.05em{\sc i\kern-.025em b}\kern-.08em
    T\kern-.1667em\lower.7ex\hbox{E}\kern-.125emX}}
\begin{document}


\title{A Hybrid Artificial Neural Network for Task Offloading in Mobile Edge Computing}

\author{\IEEEauthorblockN{Raby Hamadi, Abdullah Khanfor, \textit{Member, IEEE}, Hakim Ghazzai, \textit{Senior Member, IEEE} \\ and Yehia Massoud, \textit{Fellow, IEEE}}
{\thanks {\vspace{-0.3cm}\hrule \vspace{0.1cm}
Raby Hamadi, Hakim Ghazzai, and Yehia Massoud are with King Abdullah University of Science and Technology (KAUST), Thuwal, Makkah, KSA. (E\textendash mails: \{raby.hamadi, hakim.ghazzai, yehia.massoud\}@kaust.edu.sa).\newline
Abdullah Khanfor is with Najran University, Najran, KSA. (E\textendash mail: aikhanfor@nu.edu.sa).\newline
This paper is accepted for publication in 65th IEEE International Midwest Symposium on Circuits and Systems (MWSCAS'22), Virtual Conference, Aug. 2022. \newline \textcopyright~2022 IEEE. Personal use of this material is permitted. Permission from IEEE must be obtained for all other uses, in any current or future media, including reprinting/republishing this material for advertising or promotional purposes, creating new collective works, for resale or redistribution to servers or lists, or reuse of any copyrighted component of this work in other works.
}}
\vspace{-0.5cm}}

\maketitle

\begin{abstract}Edge Computing (EC) is about remodeling the way data is handled, processed, and delivered within a vast heterogeneous network. One of the fundamental concepts of EC is to push the data processing near the edge by exploiting front-end devices with powerful computation capabilities. Thus, limiting the use of centralized architecture, such as cloud computing, to only when it is necessary. This paper proposes a novel edge computer offloading technique that assigns computational tasks generated by devices to potential edge computers with enough computational resources. The proposed approach clusters the edge computers based on their hardware specifications. Afterwards, the tasks generated by devices will be fed to a hybrid Artificial Neural Network (ANN) model that predicts, based on these tasks, the profiles, i.e., features, of the edge computers with enough computational resources to execute them. The predicted edge computers are then assigned to the cluster they belong to so that each task is assigned to a cluster of edge computers. Finally, we choose for each task the edge computer that is expected to provide the fastest response time. The experiment results show that our proposed approach outperforms other state-of-the-art machine learning approaches using real-world IoT dataset.
\end{abstract}

\begin{IEEEkeywords}
Internet of Things (IoT), machine learning, edge computing, resource allocation, task offloading.
\end{IEEEkeywords}

\section{Introduction}

The widespread of fifth-generation (5G) broadband networks is directing to extensive deployment and continuous operation of connected devices~\cite{7414384}. 
The unprecedented scale and complexity of data generated by these devices surpass the network capacity and its infrastructure capabilities. Indeed, the employment of centralized and cloud data centers remains inefficient due to the bandwidth scarcity and latency issues in the network.

Edge Computing (EC) is a distributed computing paradigm that aims to overcome the challenge due to centralized computing~\cite{8647559,khan2019edge}.
This distributed architecture provides a more optimized and efficient solution to the process. It treats the data in terms of near computing resources such as the same geographical area, resulting in a faster response time. Consequently, EC allows a significant reduction in terms of latency and alleviates the congestion at the network back-haul and backbone levels as well as the centralized servers.

The benefits of EC is mainly observed with large and complex networks such as the Internet of Things (IoT) networks~\cite{8123913}. The IoT network operation can be significantly improved as the transmission time, latency, and bandwidth usage are optimized. For example, Microsoft took advantage of these edge infrastructures to build a system that collects videos of an open geographical zone and then processes them in real-time~\cite{8057318}. Moreover, EC provides a flexible task offloading scheme that helps maximize the lifetime of end devices' batteries~\cite{carvalho2020computation}. 


One of the most significant benefits of integrating EC with IoT networks is offloading the computing tasks to the network's edge. The adoption of embedded chips in modern IoT systems has nourished the computation capabilities of front-end devices. It is currently possible for these smart devices to perform advanced computational tasks not only local tasks but other tasks delegated by their peers. Indeed, it is now possible for IoT devices with limited computational capabilities to delegate some tasks to suitable and available edge machines instead of sending tasks back to centralized servers. The authors ~\cite{7636891} of proposed a framework to optimize the edge computer allocation using the privacy policy. Khanfor et al. proposed an intelligent community detection-based framework to allocate tasks generated by IoT devices that lack sufficient computational capabilities to edge computers within the same network~\cite{9184663}. The authors used social relationships to select trustworthy devices and employed machine learning techniques to perform end-to-end matching. A crowdsourcing-based approach to form socially connected and skilled set of devices to address the requested tasks~\cite{9345852,9446513}.
\begin{figure*}[t]
\centerline{\includegraphics[width=\textwidth]{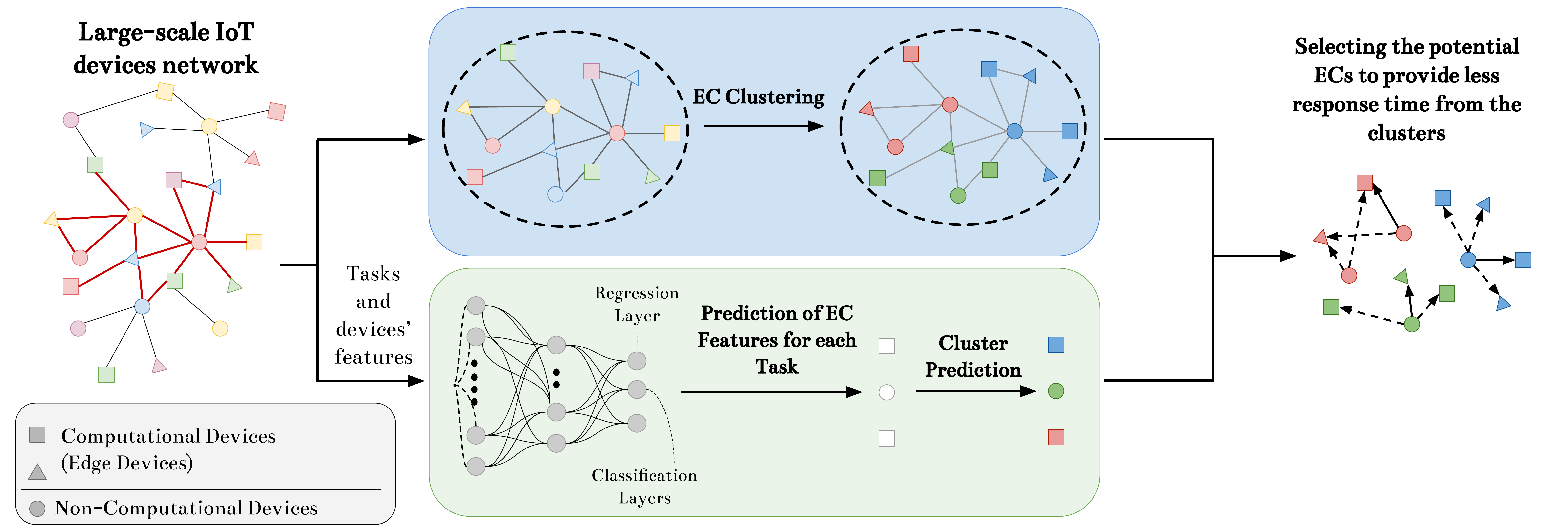}}
\caption{Proposed framework to delegate tasks to potential edge computers for computational purposes.}

\label{fig:framework}
\vspace{-.5cm}
\end{figure*}

Although this topic of recruiting potential edge computers has been investigated in literature, there is still room for improvement in the procedures used. Indeed, most of the edge computer allocation approaches are based on deterministic solutions~\cite{8336866}, which in practice remains computationally complex and cause some delays before task delegation actions occur especially in large-scale network. 
In this paper, we propose an edge computer task offloading technique based on Artificial Neural Networks (ANNs) with multiple output layers, to predict, based on different features of the tasks and the IoT devices, the features of potential edge computers that have the necessary computational power to fulfil tasks on behalf of other IoT devices that lacks of computational power. The model helps in expediting the search of edge computers in large-scale IoT networks. Extensive simulation results show the efficiency of the ANN compared to other machine learning techniques in predicting potential edge computers for different tasks.

\section{Proposed Hybrid ANN Model for Edge Computer Offloading}

In this study, we propose an offloading method for tasks that are submitted by IoT devices to be processed by an edge device within the same network. Thus, the task features of the generated task are initiated by IoT devices that lack sufficient computational power and willing to delegate this task to their capable devices (edge computers) within the same network. The potential edge computers are expected to execute these tasks on their behalf in a trustworthy and prompt manner.

\subsection{Model Architecture}
The proposed framework shown in Fig.~\ref{fig:framework} is based on a hybrid ANN with multiple output layers. Each output layer of the network is expected to predict some of the features of the edge computers. The proposed ANN architecture has two classification layers where each layer predicts the class of one of the two categorical features of the edge computers. Furthermore, the ANN has a third multi-regression layer to predict all the numerical features of the edge computers. The proposed architecture is designed to reduce the complexity of the network and avoid the usage of three different models for classification and regression. The ANN network is designed in such a way that the shallow layers are expected to detect the high-level patterns of the dataset, whereas the deeper layers, i.e., the output layers, will use these detected patterns for prediction.

In our proposed approach, we suppose to cluster all the IoT devices within the network that has computational power, i.e., edge computers, into $k$ clusters to optimize the matching process afterward. By identifying the cluster of devices in the IoT network, we are enabling the concept of fog computing where the cloud of capable devices gets closer to the edge that requires the service~\cite{zhang2017computing}. Allocating these resources and identifying them is one of the main challenges of discovering and assigning the appropriate devices in the vast IoT network. The clustering of IoT devices is supposed to occur on each timestamp $T$ to consider the dynamic availability of the IoT devices within the network.

The designed ANN is trained to recommend potential edge computer candidates to handle tasks initiated by other devices within the same network that are computationally unqualified, based on a dataset that contains the previous interactions of their peers. As shown in Fig.~\ref{fig:framework}, IoT devices with insufficient computational powers submit tasks to the network managed by a central control unit. The latter predicts potential edge computers that are able to compute these tasks. Eventually, we cluster the predicted edge computers to their potential cluster of edge computers using an unsupervised learning technique. Once we have chosen a cluster, we select the one that provides the minimum response time, denoted by $L(r,c)$, to contribute in reduce the network congestion. The total response time is calculated as follows:
\begin{equation}\label{equ:latency}
L(r,c) = CPU_\text{time} + P_\text{time},
\end{equation}
where the $CPU_\text{time}$ denotes the processing time of a task and is expressed as follows:
\begin{equation}\label{equ:cputime}
CPU_{time}= \frac{IC_r \times CPI_c }{R_c}.
\end{equation}
In~\eqref{equ:latency}, the parameter $P_\text{time}$ indicates the tasks' propagation time and is expressed as follows:
\begin{equation}\label{equ:proptime}
P_\text{time} = 2\times\frac{M_e}{R_t}.
\end{equation}
The parameter $R_c$ is the clock rate of the processor, $CPI_c$ is the number of clocks per instruction that the processor can handle, $IC_r$ is the number of instructions within the task, $M_e$ is its message size, and finally, $R_t$ is the average end-to-end transmission throughput. In this study, and without loss of generality, we consider two types of communications between the devices with regards to the distance separating them. We suppose that Device-to-Device (D2D) communication is used for short-distance communications and cellular broadband communication for long ones.
\vspace{-.3cm}
\subsection{IoT Devices Dataset}

We use an IoT dataset containing two types of devices that have computational capabilities, e.g., personal computers and smartphones, and other devices with non-computational capabilities such as weather sensors. Based on these features, we can determine edge computers, i.e., IoT devices, with sufficient computational power to process and execute a specific task initiated by other devices with limited computational resources. Thus, the dataset can be divided into two main categories: Computational and Non-Computational features.

$\bullet$ \textit{Non-computational features:}\\
The non-computational features existing in the dataset provide insights about the location of the device, its battery level, privileges (private or public), mobility (mobile or static), and its type, i.e., vehicle, tablet, mobile, laptop, etc. The dataset also includes details about the task that is generated by each IoT device, such as the task message size, the number of instructions of a task, and the task importance and priority.

$\bullet$ \textit{Computational features:} \\
Some of the IoT devices considered in this study can act as edge computers since they include the computational capability where the devices with no computational resources send these tasks to the devices with sufficient resources within the related network. The edge computers are differentiated with respect to their processors' manufacturer, processors' clock rate, clocks per instruction, installed Random Access Memory (RAM), the number of clocks per instruction of the processor installed on it, and its load that ranges between 0 and 1. 

Since there is a lack of features in the dataset for the specification of the device, we have performed a data cleansing step to deal with missing data and normalize the features. We have also generated some specifications of the device manually. For example, the availability of an edge computer is modeled as a truncated normal distribution ranging between zero and one describing the saturation of the device with respect to its RAM. To avoid getting the same values for devices having the same RAM, we add random noise using another truncated normal distribution centered on that value. 
\vspace{-.1cm}
\section{Results \& Discussions}
\label{sec:results}
All the simulations performed in this paper are based on a dataset containing information about IoT devices in the city of Santander, Spain~\cite{8580830}. The dataset comprehends a total of 16216 devices, of which 14600 from private users and 1616 from public services. In our simulations, we suppose that the clustering is performed each $T = 10$ minutes.

\subsubsection{Edge Computers Clustering}
\begin{figure}
    \centering
    \includegraphics[width=9cm]{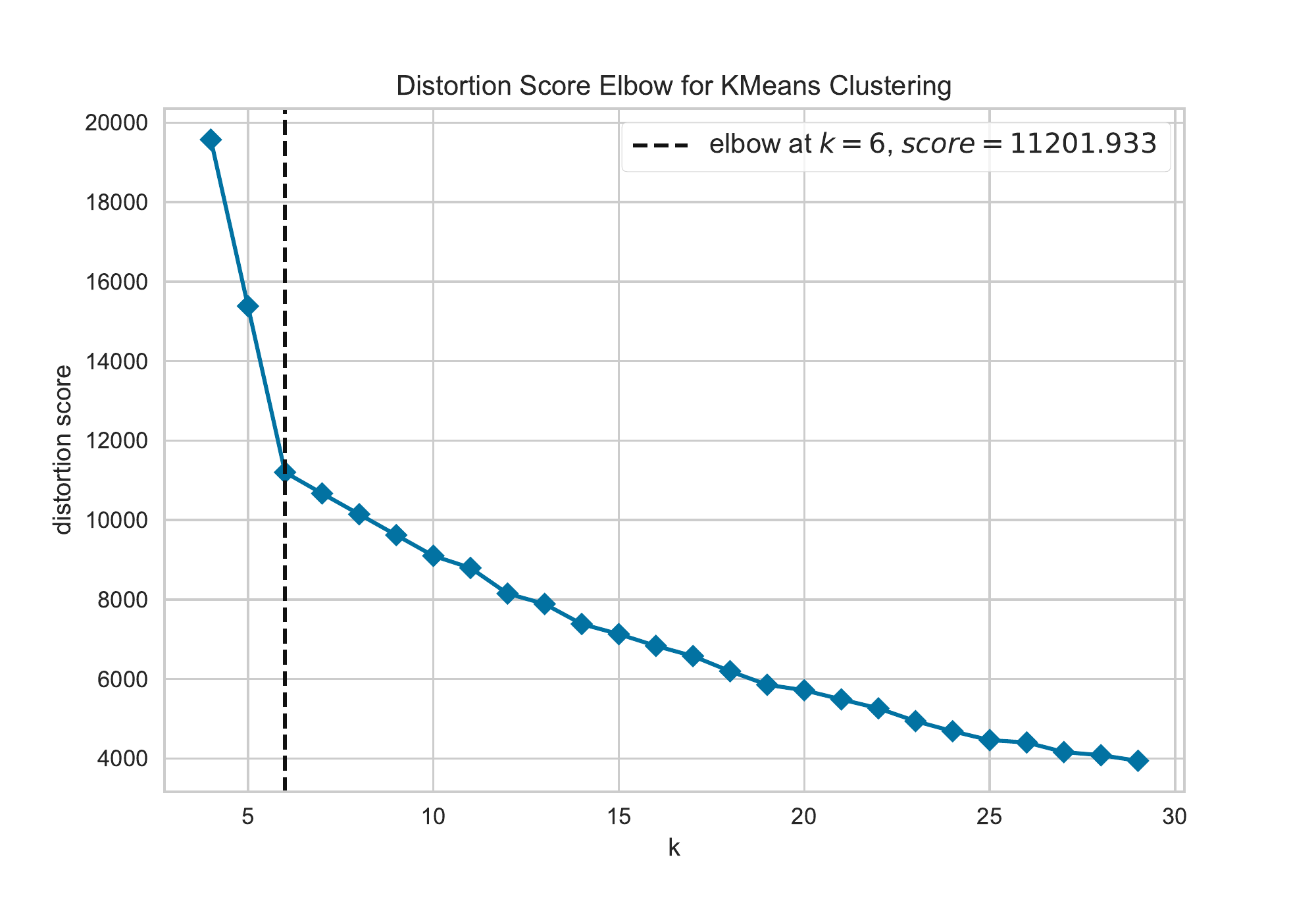}

    \caption{Distortion score variation in identifying the optimal number of clusters for K-means++ algorithm using the Elbow method.}

    \label{fig:elbow_distortion}
\end{figure}

\begin{figure}
    \centering
    \includegraphics[width=9cm]{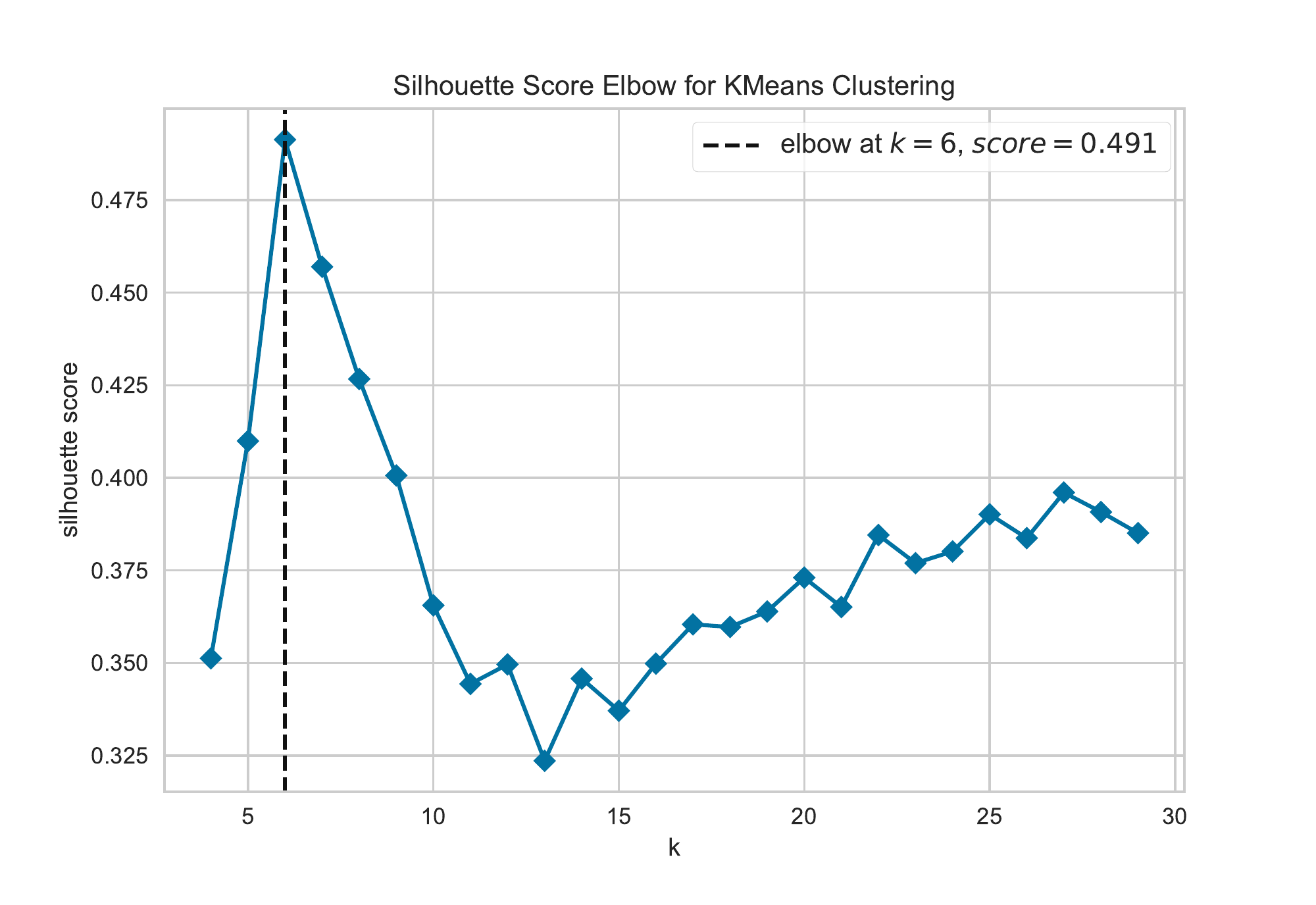}

    \caption{Silhouette score variation in identifying the optimal number of clusters for K-means++ algorithm using the Elbow method.}

    \label{fig:elbow_silhouette}
\end{figure}

In order to reduce the complexity of the framework and its execution time, we propose to cluster all the edge computers within the network. We use the unsupervised clustering technique K-means++. To select the optimal number of clusters, we apply the elbow method to select the number of clusters in our dataset using both the distortion and silhouette metrics. The distortion score is computed using the Euclidean distance to obtain the average of the squared distances from the cluster centers, whereas the silhouette uses the mean intra-cluster distance and the mean nearest-cluster distance for each sample. In this simulation, both distortion and silhouette scores gave the same optimal number of clusters $K^*=6$ as shown in Fig.~\ref{fig:elbow_distortion} and Fig.~\ref{fig:elbow_silhouette}.

\begin{figure*}[t]
\centerline{\includegraphics[width=18cm]{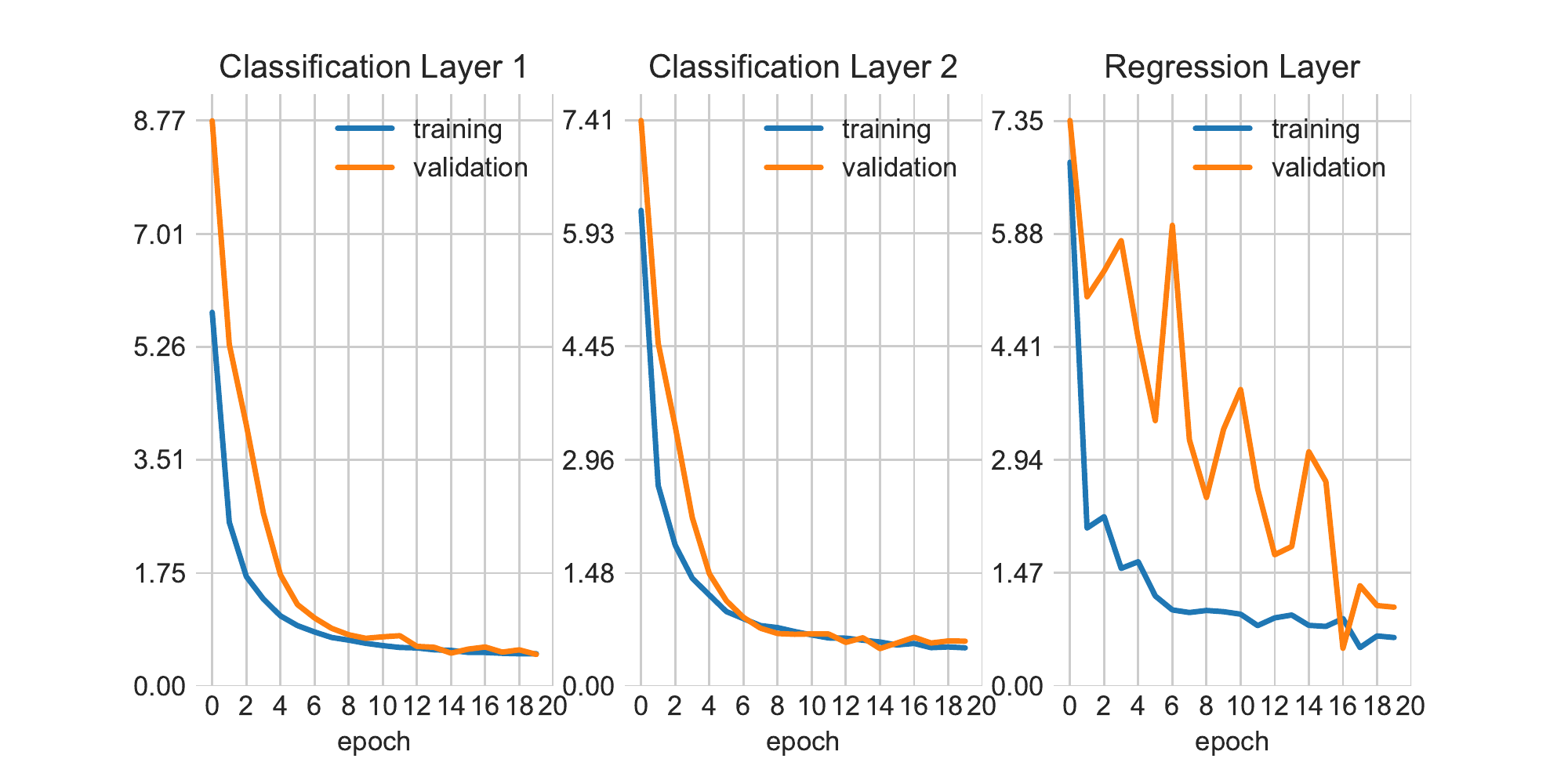}}

\caption{Hybrid Artificial Neural Network training results. Training loss vs number of epochs.}
\label{fig:train_results}
\vspace{-.5cm}
\end{figure*}

\subsubsection{Model Training}
The hybrid ANN is trained based on the previous cooperation of the IoT devices to be able to predict the potential edge computer that has enough computation power to fulfill a task initiated by another IoT device within the same network in a trustworthy and private manner. \textcolor{black}{The tasks' features i.e., the number of instructions and size as well as the requesters' features, i.e., location, type, mobility, and etc are fed to the model to predict the EC's features needed to fulfil the task.}
As our proposed architecture contains three different and separated output layers, we train each layer using a specific loss function while using the same optimization algorithm. For the classification layers, we use the categorical cross-entropy loss function and the Softmax as the activation function. As for the regression layer, we use the Mean Squared Error (MSE) as a loss function and Rectified Linear Unit (ReLU) as the activation function. To avoid overfitting, we employ early stopping and dropout layers between the dense layers of the network. The training process accomplishes after twenty epochs on batch sizes of sixteen samples. To evaluate the model performances, we use a specific evaluation metric for each output layer. We use the F1-score for the classification layers as follows:
\begin{equation}
    F1 = \frac{2 \times TP}{2\times TP + FN + FP}, 
\end{equation}
where $TP$, $FN$, and $FP$ are the number of the true positives, false negatives and false positives, respectively. 

The $\text{MSE}$ is used for the regression layers. It is defined as follows:
\begin{equation}
    \text{MSE} = \frac{1}{N}\sum_{i=1}^{N} (y-\hat{y}),
\end{equation}
where $N$, $y$, and $\hat{y}$ are the number of samples, the original output, and the predicted output, respectively. The variations of the loss functions for each layer are represented in~Fig.~\ref{fig:train_results}. After training, the hybrid model is tested on a set of unseen data. The testing results are summarized in Table~\ref{tab:train_results}.
\vspace{-.05cm}
\subsubsection{Model Validation}
To validate the performance of the proposed ANN architecture, we compare its performances in predicting suitable edge computing devices to those of other machine learning models such as decision tree regressor and gradient boosting regressor. Using the machine learning regressors, the prediction of ECs is assimilated to only a regression task where categorical features like, for example, privileges that have two values: public and private encoded as 0 and 1, will be determined threshold on the regressors' predicted value, i.e., if the predicted value is less then 0.5 we consider it as zero (public) otherwise it's a one (private). For features that have more than two categorical values, we use multiple thresholds. In this study, we use the Leave-One-Out Cross-Validation (LOOC) technique to estimate the performance of the different regression machine learning algorithms on our dataset~\cite{ref1}. Based on the results of these comparisons, we select the best machine learning algorithm, and we compare its performances with our proposed network. The results of the comparisons are depicted in Table~\ref{tab:model_comparison} where it is shown that the proposed architecture significantly outperforms other regressors. In this table, the models are compared using the MSE of the predicted values and their standard deviation $\sigma^2$. The MSE allows checking how accurate the predicted results are compared to their original values, whereas $\sigma^2$ gives insights on how close the predicted values are to the expected value. Hence, the smaller the MSE and $\sigma^2$ are, the more the predictor is accurate. The table shows the efficiency of the proposed approach compared to existing learning models with an achieved MSE of 0.568  compared to 0.865 with the gradient boosting decent algorithm.

\renewcommand{\tabcolsep}{3pt}
\begin{table}[]
\caption{Proposed network testing results}
\centering
\begin{tabular}{c c c c }
\toprule
& Classification Layer 1 & Classification Layer 2 & Regression Layer   \\
\midrule
F1-score & 0.921                & 0.903                 & -  \\
MSE  & -                 & -                 & 0.568          \\

\bottomrule
\end{tabular}
\label{tab:train_results}
\vspace{-.2cm}
\end{table}

\begin{table}[]
\caption{Expected base model performances' using the Leave-One-Out Cross-Validation procedure vs. the proposed model's performance}
\centering
\begin{tabular}{c c c}

\toprule
Model                     & MSE   & $\sigma^2$   \\
\midrule
Linear Regression (LR)         & 3.125 & 2.650     \\
Lasso Regressor (LR)                     & 5.573 & 2.300     \\
Elastic Net (EN)                & 3.733 & 3.860     \\
Decision Tree Regressor (DTR)     & 3.125 & 5.200     \\
KNeighbors Regressor (KNR)       & 3.123 & 2.720    \\
MultiOutput Regressor (MR)     & 1.536 & 1.892    \\
\textbf{Gradient Boosting Regressor (GBR)} & \textbf{0.865} & \textbf{0.568}     \\
\midrule
\textbf{Proposed Hybrid Network}   & \textbf{0.568} & \textbf{0.482}  \\
\bottomrule
\end{tabular}
\label{tab:model_comparison}
\vspace{-.5cm}
\end{table}

\vspace{-0.2cm}
\section{Conclusion}
This study proposed a deep learning-based framework that assigns a task to a suitable edge device in a mobile edge computing architecture by clustering the devices based on their profiles and then predicting the requirements to process the task and assign it to the suitable devices within the same cluster. Results show that the proposed approach outperforms the state-of-the-art models. Thus, the proposed approach can be employed as an effective tool for fast network navigation and service discovery in large-scale IoT.

\bibliography{references}
\bibliographystyle{ieeetr}
\balance

\end{document}